\documentstyle[12pt,aaspp4]{article}

\begin{document}

\title{Spectroscopy of Close Companions to QSOs and the Ages of 
Interaction-Induced Starbursts\footnotemark[1]}

\footnotetext[1]{Based on observations made with the NASA/ESA Hubble
Space Telescope, obtained from the data archive at the Space Telescope
Science Institute, which is operated by the Association of Universities
for Research in Astronomy, Inc., under NASA contract NAS 5-26555.}

\author {Gabriela Canalizo\altaffilmark{2} and Alan Stockton\altaffilmark{2}}
\affil{Institute for Astronomy, University of Hawaii, 2680 Woodlawn
 Drive, Honolulu, HI 96822}

\altaffiltext{2}{Visiting Astronomer, W.M. Keck Observatory, jointly operated
by the California Institute of Technology and the University of California.}

\begin{abstract}
We present low-resolution absorption-line spectra of three candidate close
($<$ 3\arcsec) companions to the low redshift QSOs 3CR\,323.1, PG\,1700+518 
and PKS\,2135$-$147.
The spectra were obtained with LRIS on the Keck telescopes and with the
Faint Object Spectrograph on the University of Hawaii 2.2\,m telescope. 
For 3CR\,323.1 and 
PG\,1700+518, we measure relative velocities that are consistent with an 
association between the QSOs and their companion galaxies. 
The spectral features of the companion galaxy to 3CR\,323.1 indicate a
stellar population of intermediate age ($\sim2.3$ Gyr).  In contrast, 
the spectrum of the companion object to PG\,1700+518 shows strong Balmer 
absorption lines from a relatively young stellar population, 
along with the \ion{Mg}{1}$b$ absorption feature and the 4000 \AA\ 
break from an older population.
By modeling the two stellar components of this spectrum, it is possible to 
estimate the time that has elapsed since the end of the most recent major 
starburst event:  we obtain $\sim0.1$ Gyr.  This event may have coincided 
with an interaction that triggered the QSO activity.
Finally, our spectroscopy shows conclusively that the supposed companion to 
PKS\,2135$-$147 is actually a projected Galactic G star.

\end{abstract}


\section{Introduction}

There is considerable circumstantial evidence that strong interactions 
can provide the means to feed material into a galaxy nucleus to fuel
activity in luminous QSOs
(see, e.g., Stockton 1990; Hutchings \& Neff 1992).
Such evidence includes tidal tails and
bridge-like asymmetries in QSO host galaxies, extended emission around
QSOs, and close companion galaxies. Stockton (1982)
proposed that some QSO companions might be tidally-stripped cores from 
galaxies that had interacted with the QSO host galaxies.  

Interest in QSO companions has been reawakened recently by the HST
results from a nearby luminous QSO sample.
Bahcall {\em et al.} (1997) found at least one such companion lying within
25 kpc for 13 of the 20 QSOs in their sample.   This result 
underlines the importance of companion galaxies as potential triggerers
of the QSO activity. 

However, imaging observations alone cannot distinguish true companions from
projections of other objects, so it is important to make
spectroscopic observations as well.  With precise redshift 
measurements we can calculate the relative velocities between the QSO and 
its companion(s) and determine
whether the latter are likely to be in gravitationally bound orbits. 
The relatively slow passages one obtains in bound systems appear to be 
important in bringing fresh material deeply into the center of galaxies.

Stockton (1982) measured redshifts from spectra of close companions to
Markarian\,205, 3CR\,323.1, and PKS\,2135$-$147 and found all to have 
redshifts very close to those of the respective QSOs.  He argued that all 
three were probably tidally-limited companion galaxies.
However, the redshifts of the latter two objects were 
determined from emission lines alone.   Later,
Stockton \& MacKenty (1987) found that the emission lines 
seen in their spectra come from the general extended emission around the
QSO and not specifically from the supposed companion, so the
redshifts for these two objects are no longer valid.

The main objective of this study was therefore to obtain better quality spectra 
and measure accurate redshifts for QSO companions from stellar absorption 
lines.
We remeasured the redshifts of the objects near 3CR\,323.1 and PKS\,2135$-$147.
In addition, we measured for the first time the redshift of an object recently 
found near another QSO, PG\,1700+518. 

\section{Observations and Data Reduction}

3CR\,323.1 was observed on 1991 July 19 using an image 
slicer in the Faint Object Spectrograph with a Tektronix 1024$\times$1024 
CCD on the University of Hawaii 2.2 m telescope on Mauna Kea.   The spectra 
were centered at 5500\,\AA \ and had a resolution of 2.17 \AA\,pixel$^{-1}$.
The three image slicer slits were oriented perpendicularly to the line 
joining the companion to the QSO, with the companion falling on the central
slit.   The other two slits were used to monitor the QSO light.

PG\,1700+518 and PKS\,2135$-$147 were observed on 1996 February 15
and 1996 October 13, respectively, with the Low-Resolution Imaging Spectrometer
(LRIS) on the Keck I (February) and the Keck II (October) telescopes.   
We used a 600 groove mm$^{-1}$ grating blazed 
at 5000\,\AA \ to obtain a dispersion of 1.24\,\AA \,pixel$^{-1}$ and a useful 
wavelength range of 4400--6900\,\AA\ for PG\,1700+518 and 4200--6700\,\AA \ 
for PKS\,2135$-$147.  The slit (1\arcsec\ wide, projecting to $\sim$5 pixels on
the Tektronix 2048$\times$2048 CCD) was oriented in both cases
along the line joining the companion and the QSO.    
The total exposure times were 60 minutes for 3CR\,323.1 and 40 minutes each for 
PKS\,2135$-$147 and PG\,1700+518.

The spectra were reduced with IRAF, using standard reduction procedures.
Scattered QSO light sometimes contaminated the
companion spectra, and it had to be subtracted before any 
companion spectral features were visible.
For 3CR\,323.1, no subtraction was necessary.  
For PG\,1700+518 and PKS\,2135$-$147,
the contamination amounted to 56\% and 38\%, respectively, at rest-frame
4500 \AA.
The scattered light was removed by reversing the QSO continuum
profile about its center and subtracting it from the companion
spectrum.

\section{Results and Discussion of Individual Objects}

The fields of the three QSOs are shown in Figs.\ 1 and 2 (Plates 00 and 00), 
on which objects
of interest are labelled and the slit positions indicated.
Table 1 summarizes the observed properties of the three objects.  Columns 
2 and 3 give the redshift of the QSO and the companion respectively, 
both of them obtained from our data.   In columns 4 and 5 we list the 
projected separations between the QSOs and their close companions 
in arcseconds and kiloparsecs, respectively.
In column 6 we list the velocity of the companion in the rest frame of the QSO.

\subsection{3CR 323.1}

Figure 3 shows the spectrum of the companion ($a$ in the upper panel
of Fig.\ 1) and the QSO in 
their rest frame.   While [\ion{O}{2}] and [\ion{O}{3}] emission lines 
from the extended ionized region contaminate the spectrum of the companion,
absorption features are clearly present, including the \ion{Ca}{2} H and
K lines, the G-band, and the CN feature near 3830 \AA.
    
The redshift obtained from the absorption features 
is $z = 0.2664$, which gives a velocity of $470\pm230$ km s$^{-1}$ in the QSO
reference frame defined by the narrow emission lines.  Given the uncertainty 
in both the velocity measurements and in the true systemic velocity of
the QSO host galaxy, there is a good possibility that the object is bound
to the QSO, especially if the QSO host has a fairly high mass.

Superposed on the spectrum of the companion in Fig.\ 3 is a Bruzual \& 
Charlot (1993) model for a 2.3 Gyr-old instantaneous burst stellar population
with solar metallicity.
The discrepancy in the fit is mainly due to the lower resolution in the
model, particularly evident around the \ion{Ca}{2} features. 
The intermediate-age population we find dominating the optical 
spectrum could be consistent with an aged starburst from the time of
a previous interaction between the companion and the QSO host galaxy, but
better spectra over a wider wavelength range and consideration of
more complex models would be necessary to make such a scenario compelling.

\subsection{The demise of the companion to PKS\,2135$-$147}

The object 2\arcsec\ southeast of PKS\,2135$-$147 ($a$ in the lower panel
of Fig.\ 1) was first noticed by 
Kristian (1973).
Stockton (1982) found that this object had 
a stellar or nearly stellar profile. He identified it  
as a compact companion associated with the QSO by measuring its redshift from
emission lines.   As noted previously, this redshift is invalid since the 
emission comes from an extended region surrounding the QSO, not from
the companion (Stockton \& MacKenty 1987).   Later, 
Hickson \& Hutchings (1987)
found weak evidence for the Mg\,I$b$ absorption feature at the redshift of the 
QSO in the spectrum of the companion, and Hutchings \& Neff (1992) thought 
that the object appeared extended along the radial direction towards the QSO.   

On WFPC2 images of PKS\,2135$-$147 from the HST archive, the
object has a completely stellar profile
(see Fig.\ 1).   This lack of extension 
made us suspect that the object was actually a close 
projection of a Galactic star. 
The high signal-to-noise LRIS spectrum in the top panel of Fig.\ 4 confirms this
suspicion.   Superposed on the spectrum of the object
is the spectrum of a G\,5 star, which matches closely the 
flux distribution and absorption features of the supposed companion.    

In Fig.\ 4 we also show the spectrum of galaxy $b$, 5\arcsec\ to the 
southeast of the QSO, and having a radial velocity with respect to the
QSO of $-150\pm80$ km s$^{-1}$.  This galaxy and galaxy $d$ were among 
those previously 
identified as associated galaxies (Stockton 1978).   
A much fainter galaxy on our slit 15\arcsec\ to the northwest of the 
QSO ($c$ in Figs.\ 1 and 4) has a relative velocity of $-200\pm40$
km s$^{-1}$.
Finally, we observe a strong emission region (marked by arrows and
labelled ``em'' in Fig.\ 1) beyond
companion galaxy $b$, $\sim$ 12\arcsec \ southeast of the QSO; 
its velocity with respect to the QSO is $-64\pm20$ km s$^{-1}$. 

\subsection{PG\,1700+518}

PG\,1700+518 is one of the brightest low-redshift QSOs, one of the few
low-redshift broad-absorption-line (BAL) QSOs, and a luminous IR object.
A peculiar 
characteristic of this object is an arc-like structure $\sim$ 2\arcsec \
north of the QSO (Hutchings {\em et al.} 1992; Stickel {\em et al.} 1995; 
Stockton {\em et al.} 1997; see Fig.\ 2).  It is not clear
from the imaging data whether this object is a distorted companion or 
a tidal feature from the
host galaxy, though we will generally refer to it as a 
``companion'' for convenience.  Both we and Stickel {\em et al.} (1995) find 
evidence for a fairly symmetric main component to the host galaxy, aligned
approximately E---W (see Fig.\ 2).

Figure 5 shows the spectrum of the companion as well as that of the QSO.
We measure a redshift $z=0.2929$ for the companion and $z=0.2923$ for 
the QSO, giving a relative velocity of $140\pm150$ km s$^{-1}$
in the reference frame of the QSO.  The strong [\ion{O}{2}] and [\ion{O}{3}]
lines in the spectrum come from the extended ionized gas around the
QSO and not specifically from the companion.
The apparent absorption features marked with an asterisk are 
a result of an over-subtraction of the QSO \ion{Fe}{2} emission.   
The \ion{Fe}{2} lines 
seem to be spatially variable in PG\,1700+518 (Hickson \& Hutchings
1987), leading to an uncertainty in their subtraction when we correct
for the QSO scattered light contribution.   

The companion shows a classic ``E+A'' spectrum.  The presence of strong 
Balmer absorption lines is indicative of a relatively young stellar population,
whereas the Mg\,I$b$ feature and the 4000\,\AA \ break suggest an older 
population.
We used the Bruzual \& Charlot (1993) isochrone synthesis models, with solar
metallicity and a Scalo (1986) IMF,
to model the two suggested components of the companion spectrum. 
The ages of the two isochrone stellar 
populations, determined from a $\chi^{2}$ fit to the data, are 0.09 
(+0.04,$-$0.03) Gyrs and 
12.25 Gyrs, respectively.  The younger population contributes 68\% of the 
total light at 4500 \AA\ (rest frame) and comprises $\sim6$\% of the
total luminous mass.  We caution that we are making a very simple
approximation in assuming instantaneous bursts and using only two
discrete stellar populations.  Accordingly, our best-fit model comprising
these two populations cannot be claimed to be unique, but, as shown by
the lighter curve superposed on our observed spectrum in Fig.\ 5, 
it does fit the data reasonably well.

E+A galaxies are characterized by strong Balmer absorption, fairly blue 
continuum, and weak [\ion{O}{2}] emission.   At least some
E+A galaxies seem to result from mergers (Lavery et al.\ 1992, Oegerle
et al.\ 1991), 
and Liu \& Kennicutt (1995) find a large fraction of
E+A-like galaxies in their sample of nearby merging galaxies.  Together
with the unusual morphology of the companion, this spectrum gives us
fairly strong evidence for an interaction that triggered both the QSO
activity and an extensive starburst.

\section{Age-Dating Starbursts associated with QSOs}

While determining spectral ages for stellar populations produced in a 
starburst may be the most direct approach to placing QSO hosts in an
age sequence, 
correlations among other observable parameters may ultimately give insights
into the nature of their evolution.

The spectrum of PG\,1700+518 (Fig.\ 5) shows unusually strong \ion{Fe}{2} 
emission lines for a QSO.   
Although the \ion{Fe}{2} emission in AGNs has been extensively studied
(Joly 1991 and references therein), its precise nature is still poorly
understood.   However, 
from optical spectral 
characteristics and relative positions on the far-IR $\alpha[60, 25], 
\alpha[100, 60]$ diagram, L\'\i pari (1994) suggests an evolutionary sequence 
connecting the strength of the \ion{Fe}{2} emission with the time elapsed
since a major starburst.
This scenario fits well with the proposal of 
Sanders et al.\ 1988, who have suggested an evolutionary connection between
ultraluminous IRAS galaxies (virtually all of which are interacting or
merging systems) and the classical QSO population.   

Extreme \ion{Fe}{2} objects like Mrk\,231 (Hutchings \& Neff 1987; 
L\'\i pari et al.\ 1994), close to the blackbody region in the 
FIR color--color plot, are believed to be in the last stages of merger
and in the midst of a major starburst.
Their narrow-line regions (NLR) are weak or nonexistent because the heavy 
obscuration of the QSO nucleus by dust strongly attenuates its UV
continuum.  Objects like IRAS 07598+6508 have slightly less extreme \ion{Fe}{2}
emission and a position in the FIR color--color diagram intermediate
between the black-body and the power-law regions; they are presumed to
be right at the end of the starburst stage.
Finally, strong (but not extreme) \ion{Fe}{2} emitters like
PG\,1700+518, which are still closer to the power law region of the two-color 
plot,
but which also still have very weak NLR, are in the early post-starburst stage. 
The theoretical underpinning of this model is the supposition that the
strong \ion{Fe}{2} emission arises principally in a superwind generated by
the starburst (Terlevich et al.\ 1992).  This assumption is supported by
our confirmation of the spatial variability of the \ion{Fe}{2} emission
in PG\,1700+518 (Hickson \& Hutchings 1987).

Such an evolutionary sequence is at least qualitatively consistent with 
another correlation
pointed out by Stockton \& Ridgway (1991), who noticed that the three
low-redshift QSOs that fall near the region of the FIR two-color plot
occupied by luminous FIR galaxies (Mrk\,231, Mrk\,1014, and 3CR\,48) are
also the three QSOs that show the most unambiguous evidence for tidal
tails.

Whether the arc-like object associated with PG\,1700+518 is actually
a tidal tail associated with either the host galaxy or a companion is
still uncertain, but the age we estimate since the end of the most recent 
starburst in this feature
is consistent with the evolutionary scenarios mentioned above.  
If a close interaction induced a starburst in the host galaxy as well as
in the companion 9$\times$10$^{7}$ years ago, one might well expect the system 
now to be a post-starburst object with strong \ion{Fe}{2} emission.

Thus for at least some classes of young QSOs, we have a number of
potential indicators for defining an evolutionary sequence connecting
ultraluminous IR galaxies with classical QSO population.  The strength
of the \ion{Fe}{2} emission may plausibly be related to the numbers of
Type II supernovae produced in a starburst (Terlevich et al.\ 1992).
The position in the FIR two-color plot indicates the relative balance
of thermal emission from warm dust associated with a starburst and
nonthermal emission from an AGN nucleus.  The morphology of the QSO
host galaxy and the distribution and morphologies of very close companions
may give clues to the interaction history of the immediate environment
of the QSO nucleus.  Finally, the age of the youngest host galaxy or
companion galaxy stellar population may well pinpoint the time since the
event that triggered the QSO activity.  Working out the relationships
among these various parameters will be a difficult task, but one that 
will almost certainly clarify the relationships between 
IR galaxies and QSOs.

\acknowledgments

We thank Susan Ridgway for helping with some of the observations.
This research was partially supported by NSF under grants AST92-21909 and
AST95-29078.

\newpage

\begin{deluxetable}{lccccc}
\tablewidth{0pt}
\tablecaption{Observed Properties of Close Companions}
\tablehead{\colhead{Name} & \colhead{$z_{QSO}$} & \colhead{$z_{comp}$} &
\colhead{$\Delta r$(\arcsec)} & \colhead{$\Delta r$(kpc)\tablenotemark{a}} & 
\colhead{$\Delta v$(km s$^{-1}$)}}
\startdata
3CR\,323.1      & 0.2644  &  0.2664  &  2.7  &  9.2  &      470 \nl
PKS\,2135$-$147 & 0.2005  &  0.0000  &  2.0  &  \nodata   & \nodata  \nl
PG\,1700+518    & 0.2923  &  0.2929  &  2.0  &  7.2  &      140 \nl
\enddata
\tablenotetext{a}{$H_{0}=75$ km s$^{-1}$ Mpc$^{-1}$ and $q_{0}=0.5$ assumed.}
\end{deluxetable}

\newpage

\begin{figure}
\caption{Fields of 3CR\,323.1 ({\it top panel}) and PKS\,2135$-$147 ({\it
bottom panel}), both from HST archival data.  Insets show the main regions 
of interest at lower contrast and $2.5\times$ larger scale.  The parallel
lines indicate the orientations of the slits for spectroscopy.  In both
panels, the object labelled ``$a$'' is the candidate companion galaxy.
In the PKS\,2135$-$147 field, galaxies $b$, $c$, and $d$ have been
spectroscopically confirmed to have essentially the same redshift as
the QSO.  The arrows labelled ``em'' indicate the position of a strong
emission-line component seen in our spectrum.  North is up and East to the
left in both panels.}

\caption{The field of PG\,1700+518.  The main upper panel shows an image
obtained with the University of Hawaii 2.2 m telescope (Stockton, Ridgway,
\& Kellogg 1997) in an essentially 
line-free continuum band centered at 7248 \AA\ and having a FWHM of 1260 \AA\
(5610 \AA\ and 975 \AA, respectively, in the rest frame).  The QSO has been
subtracted, using a PSF derived from a nearby star, and the white cross
indicates the QSO position.  Extensions of the parallel lines 
indicate the region covered by
the slit.  The upper-left inset shows the image before
PSF subtraction, and the lower-right inset shows a lower-contrast version
of the subtracted image at twice the scale of the main panel.  The white
oversubtracted regions are due to a slight mismatch of the profiles of
the star and the QSO; they have been suppressed in the main panel for
clarity.  The lower panel shows an $H$-band image before ({\it left})
and after ({\it right}) PSF subtraction.  The small inset shows a star
with the stretch adjusted so that the peak is the same as that of the
QSO host galaxy in the subtracted image.}

\label{3c323.1}
\caption{3CR\,323.1 spectra in rest frame. ({\em top}) Companion (heavier line)
and a 2.3 Gyr-old isochrone synthesis model. (lighter line) ({\em bottom})
QSO.   
In this and the following figures,
the spectra of the companions have been smoothed with 
Gaussian filters having $\sigma$ equal to the projected slit widths.  
The QSO spectra have not been smoothed.}
\end{figure}

\begin{figure}
\label{pks2135}
\caption{Spectra of objects associated with PKS\,2135-147.  Panels are 
identified by object labels on Fig.\ 1. ($a$) Comparison of 
PKS\,2135$-$147 ``companion'' $a$ (heavier line) and a G\,5 star (lighter line).
The spectrum of the G star has been normalized to the flux of the companion 
spectrum. 
($b$) Galaxy 5\arcsec\ southeast of the QSO, at the same redshift as the QSO, transformed to the rest frame.
($c$) Faint galaxy 15\arcsec\ northwest of the QSO, also at the same redshift.
($Q$) QSO spectrum.}

\label{pg qso+comp}
\caption{PG\,1700+518 spectra in rest frame. ({\em top}) Companion (heavier line) and isochrone synthesis model (lighter line).
({\em bottom}) QSO spectrum.}
\vspace{135mm}
\end{figure}

\end{document}